# Fluorescence-detected Fourier transform electronic spectroscopy by phase-tagged photon counting

Amr Tamimi,[1] Tiemo Landes,[2] Jonathan Lavoie,[2] Michael G. Raymer[2,*] and Andrew H. Marcus[1,*]

[1]*Department of Chemistry and Biochemistry, Center for Optical, Molecular and Quantum Science, University of Oregon, Eugene, Oregon 97403, USA*

[2]*Department of Physics, Center for Optical, Molecular and Quantum Science, University of Oregon, Eugene, Oregon 97403, USA*

\* Corresponding authors: ahmarcus@uoregon.edu, raymer@uoregon.edu

**Abstract**

Fluorescence-detected Fourier transform (FT) spectroscopy is a technique in which the relative paths of an optical interferometer are controlled to excite a material sample, and the ensuing fluorescence is detected as a function of the interferometer path delay and relative phase. A common approach to enhance the signal-to-noise ratio in these experiments is to apply a continuous phase sweep to the relative optical path, and to detect the resulting modulated fluorescence using a phase-sensitive lock-in amplifier. In many important situations, the fluorescence signal is too weak to be measured using a lock-in amplifier, so that photon counting techniques are preferred. Here we introduce an approach to low-signal fluorescence-detected FT spectroscopy, in which individual photon counts are assigned to a modulated interferometer phase ('phase-tagged photon counting,' or PTPC), and the resulting data are processed to construct optical spectra. We studied the fluorescence signals of a molecular sample excited resonantly by a pulsed coherent laser over a range of photon flux and visibility levels. We compare the performance of PTPC to standard lock-in detection methods and establish the range of signal parameters over which meaningful measurements can be carried out. We find that PTPC generally outperforms the lock-in detection method, with the dominant source of measurement uncertainty being associated with the statistics of the finite number of samples of the photon detection rate.

# 1. Introduction

Many of the most useful techniques available for studying molecular structure and dynamics are based on spectroscopy – i.e., probing the fundamental interactions between an optical field and the response function of a material sample. The principles of linear and nonlinear optical response [mu95, al87, le82] form the basis of many molecular spectroscopies such as absorption, fluorescence and Raman scattering, in addition to field correlation measurements in quantum optics. In most applications, the intensity of the light emitted from the sample (i.e. the square modulus of the signal field) is sufficiently high so that an analog optical detector can be used to produce an electrical current or voltage that is linearly proportional to the signal intensity. However, in many important situations, the signal intensity is weak and must be measured at the level of individual photons. Examples of low-flux experiments include studies of fluorescence from a single molecule or nanostructured object [ph13], or the illumination and subsequent detection of samples using quantum-entangled photon pairs [la20]. An overarching goal is often to acquire data as rapidly as possible while causing minimal optical damage to a sample. This requires extracting meaningful information from the least possible number of single-photon detection events. In this work, we establish a methodology to perform fluorescence-detected Fourier transform (FT) spectroscopy in the ultralow-signal regime, where photon counting is the preferred detection method.

In linear FT spectroscopy [da01], an optical interferometer is used to generate an optical source field resulting from interference between two optical paths. By varying the relative path delay $\tau$ or phase $\phi$, one can control the interference properties of the source field in a manner that enables extracting spectroscopic information from a molecular sample. Specifically, the source field excites the sample and the resulting signal (either in transmission or emitted fluorescence) is measured as a function of $\tau$. The recorded signal is an oscillatory $\tau$-dependent response function, which contains information about the optically induced transitions within the molecule. If the interferometer delay is stepped in increments smaller than the shortest oscillation period of the response function, the signal is said to be 'fully-sampled,' and its Fourier transform with respect to $\tau$ yields the frequency-dependent susceptibility [te06].

An alternative approach to linear FT spectroscopy, which increases data collection efficiency and the signal-to-noise ratio (SNR), is to scan the signal using a much larger step size than needed to acquire a 'fully-sampled' interferogram [te06]. This technique involves generating a reference optical beam to probe mechanical fluctuations of the interferometer at a well-defined optical frequency $\omega_R$, and to sweep continuously the phase according to $\phi(t) = 2\pi\nu t$, where $\nu$ is a 'real-time' phase-sweep frequency that is fast compared to room vibrations, but slow compared to optical frequencies. The reference beam is used to 'phase-sensitively' detect the signal interferogram (typically using a lock-in amplifier), which oscillates with respect to $\tau$ at the difference between the reference frequency $\omega_R$ and the frequencies of the optically induced transitions. The value for the reference frequency is chosen to be close to the center of the sample spectral density, such that the signal oscillates at much lower frequencies than those of the 'fully-sampled' interferogram, which is now said to be 'down-sampled.' The above described phase-sweep FT spectroscopy method has been used to sensitively detect linear and nonlinear spectroscopic signals using coherent optical sources [te06, te07, ka14, na13, br15a, br15b, gr17, ti18, br18], and more recently using quantum optical sources [la20]. Previous studies have combined phase-sweep interferometry with photon counting detection, in which a lock-in amplifier was used to determine the demodulated signal [fi06, la20, ol18]. In contrast, for high-flux measurements, high-frequency digital electronics have been used to phase-selectively detect and digitize coherent gigahertz signals [ka13, ph13, ji14, ol18, br18].

In this work, we introduce a new approach, called phase-tagged photon counting (PTPC), in which we analyzed single photon detection events to determine the linear response function and the corresponding spectrum of a resonant material sample. We used a pulsed reference laser beam, a photodetector, and high-speed electronics to monitor and assign an interferometer phase to individual photon-detection events. We compared directly the performance of the PTPC method to that of a standard lock-in amplifier. An important source of measurement uncertainty is the statistical error associated with the finite number of samples of the photon detection rate, which we simulated numerically using computer modeling. For experiments carried out under very low-flux conditions, we find that statistical sampling error is the primary source of measurement uncertainty, rather than, for example, interferometer fluctuations. Our results establish the regime of signal parameters under which useful information can be extracted under ultralow-flux conditions.

## 2. Experimental methods

### 2.1 Instrumentation

We used an apparatus similar to one implemented previously by Marcus and co-workers [te06, kr18, he19] in which a coherent source of broadband ultrafast laser pulses (center wavelength 532 nm, bandwidth FWHM 35 nm, repetition rate 140 kHz) was directed into a Mach-Zehnder interferometer (MZI, see Fig. 1). The relative path delay in the MZI (labeled $\tau$) was adjusted using a computer-controlled translation stage. Within both arms of the MZI were placed acousto-optic modulators (AOMs, labeled 1 and 2) to control the relative optical phase. The AOMs were driven continuously at fixed frequencies ($\nu_1$ and $\nu_2$) so that the relative phase of the pulses that emerged from the MZI varied in time according to $\phi(t) = 2\pi\nu_{21}t$ with $\nu_{21} = \nu_2 - \nu_1 = 5$ kHz. The laser pulses passing through the monochromator were used to monitor the net phase shift between MZI paths, and to create a reference signal for lock-in detection or for phase tagging.

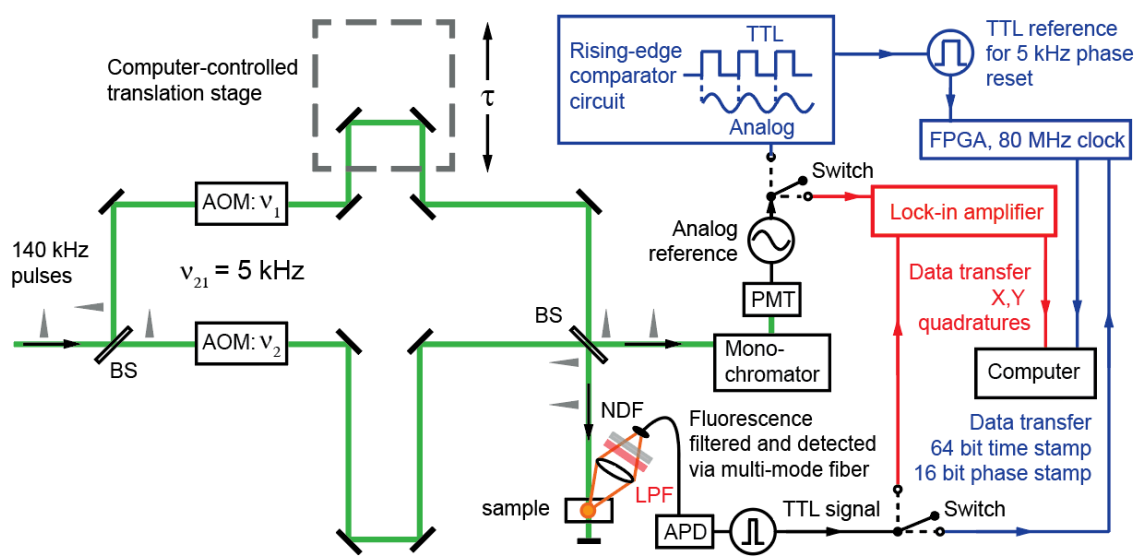

**Figure 1.** Experimental apparatus used for Fourier transform spectroscopy experiments described the text. AOM = acousto-optic modulator, BS = beamsplitter, PMT = photomultiplier tube, APD = avalanche photodiode, NDF = neutral density filter, LPF = low-pass filter, TTL = transistor-transistor-logic voltage pulses, FPGA = field-programmable gate array. Either the lock-in (red detail) or the FPGA (blue detail) data analysis electronics could be used, as selected by the two switches.

The output pulses were used to excite resonantly a sample containing an aqueous solution of a 20-base-pair segment of double-stranded DNA, in which a pair of the fluorescent dye molecules Cy3 were rigidly inserted within opposing strands of the sugar-phosphate backbones [kr18, he19]. The sample was prepared in a 1 cm quartz cuvette at 1 $\mu$M concentration in a standard aqueous buffer of 10 mM TRIS, 100 mM NaCl, and 7 mM $MgCl_2$. The optical absorbance spectrum of the Cy3 dimer DNA system is well described using a simple Holstein Hamiltonian model, in which the two-electronic-level transitions of the Cy3 monomers are each coupled to a single harmonic vibration, and the monomers are themselves coupled to each other through an electrostatic transition charge interaction [kr18, he19].

Fluorescence was detected at a 45° angle of incidence from the front face of the sample cuvette using a 5 cm collection lens and a fiber-coupled avalanche photodiode (APD, Laser Components, COUNT-10B-FC, dark count rate ~ 10 Hz). The detection geometry was chosen to minimize the amount of excitation light scattered into the detection pathway. Scattered light was further removed using a 615 nm long-pass filter (LPF, Chroma, HQ615LP). The fluorescence rate (i.e. the signal flux) was attenuated to single-photon count levels using a series of absorptive neutral density filters (NDFs, Thorlabs NE10A). The APD module creates a stream of TTL voltage pulses, which was sent to a lock-in amplifier (Stanford Research Systems SR830, see Fig. 1, red detail), or alternatively (see blue detail) to a field-programmable gate array (FPGA, National Instruments, PXIe-7971R FlexRIO FPGA Module Kintex-7 K325T, NI PXIe-1073 Integrated MXIe chassis, NI 5732 14-bit, 80 MHz adapter module). In the following sections, we describe these two detection methods in further detail.

For the discussions that follow, it is useful to consider the 'high-flux' versus 'low-flux' signal regimes. In the high-flux regime, many photons are incident at a detector during its response time. A high-flux signal is typically monitored using a linear-response detector (e.g., a photomultiplier tube with 1 ns response time) to produce a current or voltage proportional to the running-average flux [ha07]. As in previous studies using cw excitation [ph13, fi06], we regard the high-flux-limiting signal as equivalent to the probability to detect a single photon under low-flux conditions. In the photon counting (low-flux) experiments of our current study, each photon detection event results in a single TTL voltage pulse, without the ability to resolve multiple simultaneous events. In order to avoid loss of signal information due to saturation of the APD [oc84], we attenuated the signal so that the photon rate incident at the APD was well below our laser repetition rate (140 kHz pulsed laser source).

## 2.2 Fourier transform spectroscopy with phase-sweeping and lock-in detection

For a fixed MZI time delay $\tau$, the ensemble-averaged photon count rate can be described as a sum of two terms, which exhibit unique dependences on the phase sweep $\phi(t)$ [te06]

$$A(\phi, \tau) = A_{\text{bkgd}} + A_{\text{lin}}(\phi, \tau) \tag{1a}$$

$$= 2\sum_n |\alpha(\omega_{ng})|^2 |\mu_{ng}|^2 + 2\text{Re}\sum_n |\alpha(\omega_{ng})|^2 |\mu_{ng}|^2 \exp\{i[\phi(t) - \omega_{ng}\tau - \theta_{\text{lin}}(\omega_{ng}, \tau)]\} \tag{1b}$$

The first term on the right-hand side of Eq. (1a), $A_{\text{bkgd}}$, represents a constant background with respect to the phase sweep, and arises from the independent action of the two laser pulses. The background component may include additional contributions such as stray light at the detector, or electronic noise. The second term, $A_{\text{lin}}$, varies at the phase-sweep rate $\nu_{21} = \phi(t)/2\pi t$ and represents the 'interference contribution' to the linear-response excited-state population, which results from the collective action of the two laser pulses [te06]. Equation (1b) relates the linear-response excited-state population to model parameters that characterize the laser spectrum and the molecular system and assumes the absence of background noise and perfect alignment of the interferometer. In Eq. (1b), the sum is carried over all molecular excited-state levels (labeled $n$), $\mu_{ng}$ is the transition dipole matrix element that couples the ground and $n$th excited states, $\omega_{ng} = \widetilde{\omega}_{ng} - i\gamma_{ng}$ is the complex-valued optical transition frequency, $|\alpha(\omega)|^2$ is the intensity of the laser pulse spectrum, and $\theta_{\text{lin}}(\omega_{ng}, \tau)$ is the phase associated with the transition for the interferometer delay $\tau$ [te06]. The complex transition frequency accounts for the optical dephasing rate $\gamma_{ng}$ (i.e., the homogeneous line half width), and $\widetilde{\omega}_{ng} = (\varepsilon_n - \varepsilon_g)/\hbar$ is the optical resonance frequency.

Higher-order signal contributions to the excited-state population may also be present at multiples of the phase-sweep rate $n\nu_{21}$ (with $n$ = 2, 3, …). Such terms represent interference contributions to the nonlinear excited-state population, which include stimulated emission and excited-state absorption processes [mu95, pe12, br15a, br15b, br17, ka16]. In the current work, we consider only the linear signal response.

In all of our experiments, we implemented the 'down-sampling' technique developed previously for low-signal FT spectroscopy [te06, te07], and recently applied to entangled-photon-pair interferometry [la20], by monitoring the beam emerging from the second output port of the MZI. This 'reference' beam was spectrally filtered using a monochromator set to the wavelength $\lambda_R = 2\pi c/\omega_R$ = 515 nm, and detected using a photomultiplier tube (PMT), which produced a photocurrent signal proportional to $\cos[\phi(t) - \omega_R \tau]$.

The phase-sweeping approach provides a useful strategy to sensitively and separately determine the fluorescence signal amplitudes $A_{\text{bkgd}}$ and $A_{\text{lin}}$. The relative magnitudes of $A_{\text{bkgd}}$ and $A_{\text{lin}}$ will differ for various types of samples and are also influenced by the spatial and spectral overlap of the optical beams emerging from the MZI. To characterize the nature of a given signal, we define the flux $f$ and the visibility $v$. The flux is the average signal integrated over a full modulation cycle, $f \equiv \frac{1}{2}(A_{\max} + A_{\min}) = A_{\text{bkgd}}$, which is equal to the (phase-independent) background amplitude. The visibility is the relative signal modulation $v \equiv (A_{\max} - A_{\min})/(A_{\max} + A_{\min}) = A_{\text{lin}}/A_{\text{bkgd}}$, which is the ratio of the linear signal amplitude to that of the background. We may thus rewrite Eq. (1):

$$A(\phi, \tau) = f \left\{ 1 + \frac{2}{f} \sum_n |\alpha(\omega_{ng})|^2 |\mu_{ng}|^2 \cos[\phi(t) - \omega_{ng}\tau - \theta_{\text{lin}}(\omega_{ng}, \tau)] \right\} \quad (2a)$$

$$= f\{1 + v(\tau)\cos[\phi(t) - \psi(\tau)]\} \quad (2b)$$

where we have emphasized in Eq. (2b) that the signal visibility $v(\tau)$ and phase $\psi(\tau)$ are functions of the interferometer delay $\tau$.

Equation (2) represents the high-flux-limiting signal that would result if a linear detector were used in place of an APD. It describes the probability that the APD detects a single photon at time $t$ as a function of the experimental control parameters $\phi$ and $\tau$. We next review briefly the operation of the lock-in amplifier, which in the conventional method demodulates the analog signal described by Eq. (2). The lock-in multiplies the signal by cosine and sine waveforms derived from the reference and performs a low-pass filtering operation, which averages the signal phase over multiple modulation cycles during the lock-in time period $T_{LI}$ (= 100 ms). The lock-in thus provides the in-phase (cosine transform) and in-quadrature (sine transform) components of the linear signal response, according to

$$X_{\text{lin}}(\tau) = \frac{2}{T_{LI}} \int_0^\infty A[\phi(t), \tau] \cos[\phi(t) - \omega_R \tau] e^{-t/T_{LI}} dt \tag{3a}$$

$$= 2 \sum_n |\alpha(\omega_{ng})|^2 |\mu_{ng}|^2 \cos[(\omega_{ng} - \omega_R)\tau + \theta_{\text{lin}}(\omega_{ng}, \tau)] \tag{3b}$$

$$= fv(\tau) \cos[\psi(\tau) - \omega_R \tau] \tag{3c}$$

$$Y_{\text{lin}}(\tau) = -\frac{2}{T_{LI}} \int_0^\infty A[\phi(t), \tau] \sin[\phi(t) - \omega_R \tau] e^{-t/T_{LI}} dt \tag{4a}$$

$$= -2 \sum_n |\alpha(\omega_{ng})|^2 |\mu_{ng}|^2 \sin[(\omega_{ng} - \omega_R)\tau + \theta_{\text{lin}}(\omega_{ng}, \tau)] \tag{4b}$$

$$= -fv(\tau) \sin[\psi(\tau) - \omega_R \tau] \tag{4c}$$

From Eqs. (3) and (4), we see that the down-sampled interferograms vary as a function of delay at the frequencies $(\omega_{ng} - \omega_R)$, which are much lower than the optical frequencies of either the laser or the peak molecular absorbance. Thus, the down-sampling method requires fewer MZI-delay steps to fully characterize the information contained by the linear response. We further note that the down-sampled interferogram is more stable with respect to instabilities of the MZI path delay and phase than the corresponding fully sampled interferogram.

The linear signal components produced by the lock-in, as described by Eqs. (3c) and (4c), can be combined to construct the complex-valued interferogram

$$Z_{\text{lin}}(\tau) = X_{\text{lin}}(\tau) - iY_{\text{lin}}(\tau) \tag{5a}$$

$$= fv(\tau)e^{i[\psi(\tau)-\omega_R\tau]} \tag{5b}$$

with amplitude $|Z_{\text{lin}}(\tau)| = fv(\tau)$ and down-sampled phase $\psi(\tau) - \omega_R\tau$. By performing a Fourier transform of the signal interferogram with respect to the interferometer delay, we may obtain the complex-valued overlap spectrum $\hat{Z}_{\text{lin}}(\omega)$, which is the product of the laser spectrum and the molecular susceptibility (see below).

$$\hat{Z}_{\text{lin}}(\omega) = \int_0^\infty Z_{\text{lin}}(\tau)e^{i\omega\tau}d\tau \tag{6a}$$

$$= \int_0^\infty fv(\tau)e^{i\psi(\tau)}e^{i(\omega-\omega_R)\tau}d\tau \tag{6b}$$

$$= \sum_n |\alpha(\omega_{ng})|^2 |\mu_{ng}|^2 \int_0^\infty e^{-i[(\omega_{ng}-\omega_R)\tau+\theta_{\text{lin}}(\omega_{ng},\tau)]}e^{i\omega\tau}d\tau \tag{6c}$$

Equations (6a) and (6b) refer to experimental data processing, while Eq. (6c) describes the laser-molecule overlap spectrum in terms of the model Hamiltonian parameters. Equation (6c) can be rewritten compactly as [te06]:

$$\hat{Z}_{\text{lin}}(\omega - \omega_R) = -i|\alpha(\omega)|^2 \hat{\chi}^{(+)}(\omega) \tag{7}$$

where $\hat{\chi}^{(+)}(\omega) = i\sum_n |\mu_{ng}|^2 \int_0^\infty e^{-i[\omega_{ng}\tau+\theta_{\text{lin}}(\omega_{ng},\tau)]}e^{i\omega\tau}d\tau$ is the positive-frequency component of the complex-valued molecular susceptibility.

We note that the mean signal flux $f = A_\text{bkgd}$ can be determined by omitting the lock-in amplifier from the detection scheme shown in Fig. 1 and measuring directly the phase-averaged signal for a given MZI path delay.

### 2.3 FT spectroscopy by phase-tagged photon counting (PTPC)

In the previous section, we reviewed how down-sampling with lock-in detection can improve the signal-to-noise ratio (SNR) and reduce the number of MZI steps needed to perform Fourier transform spectroscopy in the high-flux regime [te06, fi06, ka16, br18, la20]. However, for photon counting experiments at low-flux levels, the use of a lock-in amplifier cannot fully retain the information available from the TTL pulse stream, each pulse of which is convolved with the lock-in instrument response function, as described by Eqs. (3a) and (4a).

We measured the *X*- and *Y*-quadrature signals from the TTL pulse stream using two separate procedures. In the first procedure, we sent the TTL signal to the lock-in amplifier, while using the analog signal from the PMT-monitored monochromator as the lock-in phase-reference (see Fig. 1, red detail). In the second procedure, we applied our newly implemented PTPC method using femtosecond laser pulses, which has similarities to an approach developed previously for CW excitation [fi06, ph13]. In our PTPC experiments, we used a field-programmable gate array (FPGA) to discretize the phase of a given phase-sweep cycle into a set of *m* 'phase bins,' which were numbered and incrementally advanced using an 80 MHz digital counter (Fig. 1, blue detail). The FPGA is a manually reconfigurable integrated circuit that contains hardware-enabled signal-processing algorithms. We first converted the 5 kHz analog reference waveform into a logical square wave using a stand-alone comparator circuit. The resulting reference square wave was used to trigger an 80 MHz phase counter (16-bit width), and to reset the counter at the 5 kHz phase-sweep frequency. The counter reset automatically maintained synchronicity with the phase-sweep cycle in the presence of phase fluctuations of the interferometer reference. The average number of phase bins during a phase-sweep cycle was $m$ = 80 MHz / 5 kHz = 16,000 bins, and the phase bin interval was $\Delta\phi = 360°/16{,}000 \text{ bins} = 0.0225°$ bin$^{-1}$. Thus, during each phase-sweep cycle the counter incremented the phase bin value $\phi_j = j\Delta\phi$ (with $j$ = 0, 1, …, $m$ – 1) at the 80 MHz clock speed. Individual photon detection events were assigned their respective 16-bit phase bin values $\phi_j$, which were streamed to computer memory. We also used a second 80 MHz counter (with 64-bit width) to assign a 'time stamp' to each TTL detection event. Although we did not make use of the time stamp in this current work, we note that such information is generally useful for studies of non-stationary systems, such as single-molecule emitters.

To conceptualize how the TTL pulse stream is related to the high-flux signal rate $A(\phi, \tau)$, we can create a histogram of the PTPC detection events with respect to $\phi$. In Fig. 2, we show an example histogram of a relatively large number of PTPC detection events ($N = 120,000$), which we obtained from the Cy3 dimer DNA sample using the apparatus shown in Fig. 1 with MZI path delay $\tau = 0$. Superimposed with these data is the theoretical ensemble-averaged count rate $A(\phi, \tau)$ [Eq. (2b), black curve] with flux $f = 2,000$ s$^{-1}$, visibility $v = 0.75$ and phase $\psi = 0°$ (corresponding to a delay $\tau = 0$). For this relatively large number of detection events, the $\phi$-dependance of the PTPC histogram exhibits accurately the expected sinusoidal shape of $A(\phi, \tau)$.

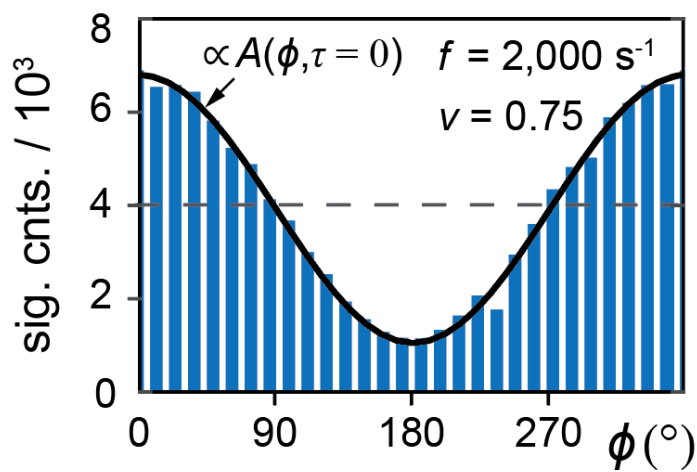

**Figure 2.** Measured phase histograms of the phase-tagged photon count (PTPC) rate from the Cy3 dimer DNA construct for data integration time window 60 s, and MZI path delay $\tau = 0$. The signal conditions were flux $f = 2,000$ s$^{-1}$ and visibility $v = 0.75$. For the purpose of visualization, the histogram was constructed by re-binning the native 16,000 bin histogram into 29 bins, so that the resolution shown is course-grained to $\sim 12.4°$ bin$^{-1}$. The horizontal dashed line indicates the average number of counts per bin.

The favorable comparison between the ensemble-averaged event histogram and Eq. (2b) shown in Fig. 2 suggests the following model to reconstruct the *mean* complex-valued interferogram $\bar{Z}(\tau) = \bar{X}(\tau) + i\bar{Y}(\tau)$ from the PTPC signal. The following method is superior to simply fitting histogrammed data to a cosine function. As we discussed in the previous section, we regard $A(\phi, \tau)$ as a continuously varying probability to observe a single photon at time $t$ for a given MZI delay $\tau$ and phase $\phi(t) = 2\pi\nu_{21}t$. The PTPC observable is the phase of the interferometer at each detection event, which we interpret as the conditional probability that the interferometer will have phase $\phi(t)$ when a photon is detected. We calculate this probability from $A(\phi, \tau)$ according to:

$$P(\phi|\text{event}) = \frac{1}{2\pi f} A(\phi, \tau) = \frac{1}{2\pi}\{1 + v(\tau)\cos[\phi - \psi(\tau)]\} \tag{8}$$

For a given MZI delay $\tau$, we consider each TTL detection event as a Dirac delta function $\delta(\phi - \phi_j)$, where the $\phi_j$ are random variables distributed according to Eq (8). We define the phase-dependent photon rate from a set of $N$ detection events within a fixed time window $T_{PT}$ as the discrete sum: $A^{PT}(\phi, \tau) \equiv \frac{1}{T_{PT}}\sum_{j=1}^{N}\delta(\phi - \phi_j)$, where $PT$ refers to 'phase-tagged.' A straightforward way to isolate the mean linear signal component from background is to calculate the first term of the Fourier series expansion of the PTPC signal with respect to $\phi$. This is mathematically equivalent to summing over the $N$ photon phase factors as in the second equality of Eq. (9b):

$$\bar{Z}_{\text{lin}}^{PT}(\tau) \equiv \frac{1}{\pi}\int_0^{2\pi} A^{PT}(\phi, \tau)e^{-i\phi}d\phi \tag{9a}$$

$$= \frac{1}{\pi T_{PT}}\int_0^{2\pi}\left[\sum_{j=1}^{N}\delta(\phi - \phi_j)\right]e^{-i\phi}d\phi = \frac{1}{\pi T_{PT}}\sum_{j=1}^{N}e^{-i\phi_j} \tag{9b}$$

$$= \bar{f}^{PT}\bar{v}^{PT}(\tau)e^{i[\bar{\psi}^{PT}(\tau) - \omega_R\tau]} \tag{9c}$$

The mean quadratures of the linear PTPC signal may be obtained according to:

$$\bar{Z}^{PT}_{\text{lin}}(\tau) = \bar{X}^{PT}_{\text{lin}}(\tau) - i\bar{Y}^{PT}_{\text{lin}}(\tau) \tag{10a}$$

$$\equiv \frac{1}{\pi T_{PT}} \left[ \sum_{j=1}^{N} \cos(\phi_j) - i \sum_{j=1}^{N} \sin(\phi_j) \right] \tag{10b}$$

The mean background signal component (i.e. the mean flux during the time window $T_{\text{PT}}$) may be obtained by averaging the PTPC rate over the phase variable:

$$\bar{f}^{PT} \equiv A_{\text{bkgd}} = \frac{1}{2\pi} \int_0^{2\pi} A^{PT}(\phi,\tau) d\phi \tag{11a}$$

$$= \frac{1}{2\pi T_{PT}} \int_0^{2\pi} \left[ \sum_{j=1}^{N} \delta(\phi - \phi_j) \right] d\phi = N/T_{PT} \tag{11b}$$

Equation (9) describes the complex-valued mean linear response function $\bar{Z}^{PT}_{\text{lin}}$ obtained by averaging the phase factor $e^{-i\phi}$ over the discrete set of $N$ detection events given by $A^{PT}(\phi,\tau)$, which were measured during the time window $T_{PT}$. Under low-flux conditions, the mean linear signal $\bar{Z}^{PT}_{\text{lin}}$ is characterized by the mean phase $\bar{\psi}^{PT}$ and mean amplitude $\bar{f}^{PT}\bar{v}^{PT}$, or alternatively by the mean quadratures $\bar{X}^{PT}_{\text{lin}}$ and $\bar{Y}^{PT}_{\text{lin}}$. In general, these signals are subject to statistical uncertainties associated with the finite number sampling of the probability distribution given by Eq. (8). As we discuss in the following sections, these uncertainties diminish with increasing flux levels and approach zero in the limit $N \to \infty$.

Although it is not advantageous to create statistical histograms of the PTPC rate $A^{PT}(\phi,\tau)$ to compute $\bar{Z}^{PT}_{\text{lin}}$, as described by Eq. (9) above, viewing the histograms can provide insights about the information available from the PTPC data stream. In Fig. 3, we present further examples of such histograms obtained from the Cy3 dimer DNA sample with MZI path delay $\tau = 0$ for four different signal conditions. For these measurements, the flux was controlled by inserting various neutral density filters into the detection path, while the visibility was adjusted by introducing stray background light into the path. Data acquisition using the FPGA was carried out for several seconds, and the mean visibility $\bar{v}^{PT}$ and flux $\bar{f}^{PT}$ were calculated according to Eqs. (9) and (11). The resulting values are shown in the insets of each panel Fig. 3, and plots of the corresponding ensemble-averaged count rates $A(\phi, \tau)$ [Eq. (2b), black curves] are shown superimposed with the histograms. In the limit of large sample number $N$, each histogram should converge to the cosine probability distribution of Eq. 8. These data indicate that for even relatively large values of the visibility, the phase-dependent modulation of the discrete photon count rate $A^{PT}(\phi, \tau)$ is increasingly difficult to perceive for very low-flux levels. In the following section, we discuss in further detail how the statistical uncertainty of these measurements is related to the signal parameters.

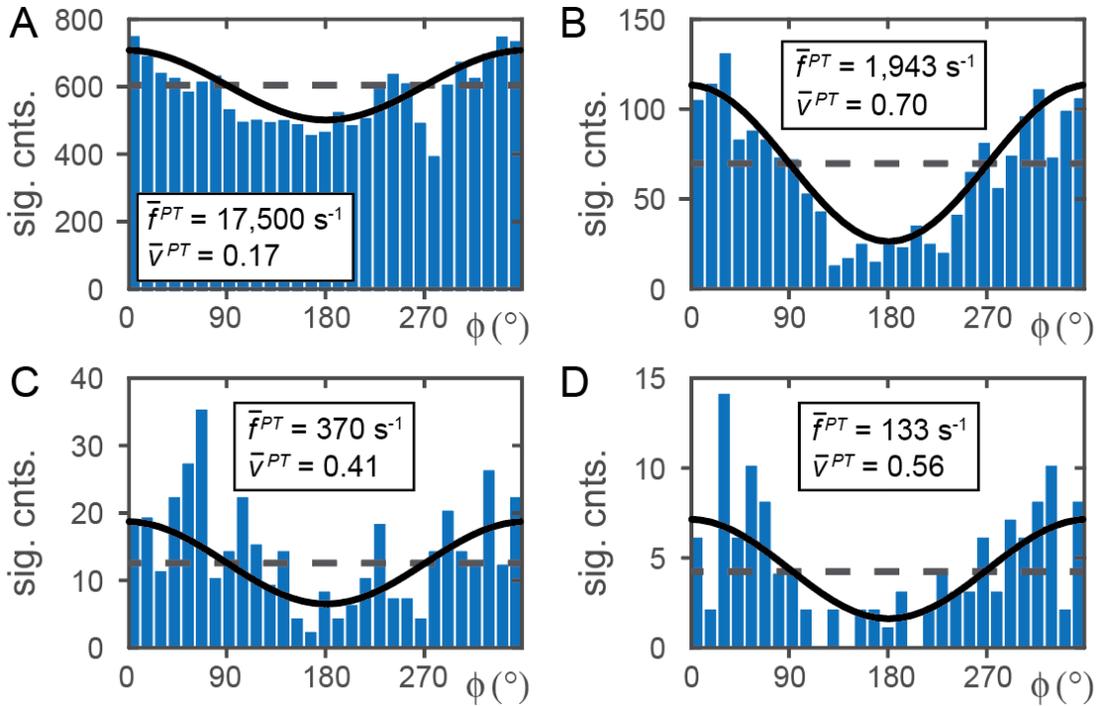

**Figure 3.** Measured phase histograms of the PTPC rate measured from the Cy3 dimer DNA construct for data integration time window $T_{PT} = 1$ s and MZI path delay $\tau = 0$. The signal conditions vary as indicated according to flux and visibility: (**A**) $\bar{f}^{PT} = 17{,}500$ s$^{-1}$ and $\bar{v}^{PT} = 0.17$, (**B**) $\bar{f}^{PT} = 1{,}943$ s$^{-1}$ and $\bar{v}^{PT} = 0.70$, (**C**) $\bar{f}^{PT} = 370$ s$^{-1}$ and $\bar{v}^{PT} = 0.41$ and (**D**) $\bar{f}^{PT} = 133$ s$^{-1}$ and $\bar{v}^{PT} = 0.57$. The histograms were re-binned as described in the caption of Fig. 2. In each panel, the horizontal dashed line indicates the average counts per bin, and the solid curves are fitted cosine functions for visualization.

## 2.4 Standard error (SE) and signal-to-noise ratio (SNR) for lock-in detection and phase-tagged photon counting (PTPC)

One of the principle findings of this work is that the sparse and statistical sampling under low-flux conditions of the phase-dependent signal $A(\phi, \tau)$ is the dominant contribution to the uncertainty of these measurements. This result is satisfactory, as it indicates that we are extracting the maximum amount of information available from the signal photon stream. In the following sections, we consider the evaluation of measurement uncertainty for both the lock-in detection and PTPC methods. Understanding the origins of the uncertainty will permit us to compare the performance of the two methods under varying signal conditions.

In accordance with the central limit theorem, we expect the outcomes of $N$ statistically independent measurements of the quantity $\alpha$ ($\alpha_1, \alpha_2, \ldots, \alpha_N$) to be distributed as a Gaussian with mean value $\bar{\alpha} = \frac{1}{N}\sum_{i=1}^{N}\alpha_i$ and standard deviation $\sigma_\alpha = \left[\frac{1}{N-1}\sum_{i=1}^{N}(\alpha_i - \bar{\alpha})^2\right]^{1/2}$ [gr04]. We thus estimate the signal-to-noise ratio (SNR) for a set of $N$ independent measurements performed over a specified time window as $\bar{\alpha}/\sigma_\alpha$, which scales as $\sqrt{N}$. We also calculate the rate-normalized standard error (SE) as $\sigma_\alpha/N$, which scales as $1/\sqrt{N}$. To compare the performances of lock-in detection versus PTPC methods, we carried out a series of measurements under identical experimental conditions.

For our measurements using the lock-in amplifier, we consider the instrumental time window over which a single data point could be acquired. The in-phase ($X_{\text{lin}}$) and in-quadrature ($Y_{\text{lin}}$) signals are represented by the convolution integrals given by Eqs. (3) and (4), which include an exponential decay with time constant $T_{LI}$. Because the exponential term decays to zero for $t \gg T_{LI}$, we adopted the convention of equating the lock-in time window with the so-called 'settling time,' which is $\sim 10\, T_{LI}$ when the lock-in filter slope was set to 24 dB/octave. In practice, for each step of the interferometer delay we recorded the quadrature signals over a dwell period of $10\, T_{LI}$ to determine the response functions $X_{\text{lin}}(\tau)$ and $Y_{\text{lin}}(\tau)$. To determine the measurement uncertainties, we set the interferometer delay to zero, for which case the visibility was maximized, and recorded the quadrature values over an extended period $n \times 10\, T_{LI}$. We thus determined a set of $n$ statistically independent measurements by dividing the trajectories into equally spaced segments of duration $10\, T_{LI}$. From the set of $n$ measurements, we determined the mean quadrature signals, $\bar{X}_{\text{lin}}(0)$ and $\bar{Y}_{\text{lin}}(0)$, and their corresponding standard deviations, $\sigma_{X_{\text{lin}}}$ and $\sigma_{Y_{\text{lin}}}$.[1] In addition, we transformed the set of $n$ quadrature signals into a set of $n$ complex-valued signals $Z_{\text{lin}} = X_{\text{lin}} - iY_{\text{lin}} = fve^{i[\psi - \omega_R \tau]}$, from which we evaluated the mean and standard deviation of the amplitude $|Z_{\text{lin}}| = fv$ and the phase $\psi$.

In our PTPC experiments, for each interferometer delay we collected $N$ photons and their respective phase assignments during the time window $T_{PT}$. In contrast to the lock-in amplifier measurements, which consist of instrumentally processed $X_{\text{lin}}$- and $Y_{\text{lin}}$-quadrature values, the PTPC measurements consist of pre-processed information about the set of $N$ photon detection events and their corresponding MZI phases. From the set of $N$ measurements, we calculated the mean complex-valued signal $\bar{Z}_{\text{lin}}^{PT}$ according to Eq. (9), or alternatively the mean quadrature signals $\bar{X}_{\text{lin}}^{PT}$ and $\bar{Y}_{\text{lin}}^{PT}$ according to Eq. (10). Each such data set is a finite sample of the normalized probability distribution described by Eq. (8). To take full advantage of the information provided by the finite PTPC data sets, we implemented a 'bootstrapping' procedure [ef94], in which we considered the $N$ sampled data points as a representative population, to determine the standard deviation $\sigma_\alpha$ of the mean signal $\bar{\alpha}$ $(= \bar{X}_{lin}^{PT}, \bar{Y}_{lin}^{PT}, |\bar{Z}_{\text{lin}}^{PT}|, \text{ and } \bar{\psi}^{PT})$. By assuming that the $N$ sampled data points can be treated as a representative population, we randomly resampled (with replacement) this population to generate $B$ 'bootstrap samples,' each of which contained $N$ elements. From each bootstrap sample we calculated a 'bootstrap estimate' of the sample statistic $\bar{\bar{\alpha}}_i$ ($i = 1,2, ..., B$), from which we calculated the 'bootstrap standard error' $\sigma_{\bar{\alpha}} = \left[\frac{1}{B-1}\sum_{i=1}^{B}(\bar{\bar{\alpha}}_i - \bar{\alpha})^2\right]^{1/2}$.

We determined analytical expressions for the expected mean and variance of the PTPC quadrature signals as a function of the specified parameters of the probability distribution $P(\phi)$ [Eq. (8)], which are the flux $f = N/T_{PT}$, the visibility $v$, and the phase $\psi$. For example, during the integration time window $T_{PT}$, the expected mean $X$-quadrature signal is calculated according to: $\bar{X}_{lin}^{PT} = N\langle\cos\phi\rangle_\phi = \frac{N}{\pi}\int_0^{2\pi}\cos\phi[1 + v\cos(\phi - \psi)]d\phi = Nv\cos\psi$. A similar calculation for the mean $Y$-quadrature signal yields: $\bar{Y}_{lin}^{PT} = Nv \sin \psi$. We further determined the standard deviations of the quadrature signals.

---

[1] We note that the above calculation of the uncertainty corresponds to a single settling time period 10 $T_{LI}$, and not the average over the extended period $n \times 10\ T_{LI}$. Our calculation thus represents the standard error of the measurement during the settling period 10 $T_{LI}$, which need not be further divided by the factor $\sqrt{n}$.

$$\sigma_{X_{lin}^{PT}} = \sqrt{\langle (X_{lin}^{PT} - \bar{X}_{lin}^{PT})^2 \rangle_\phi} = \sqrt{2N\left(1 - \frac{v^2}{2}\cos^2\psi\right)} \quad (12a)$$

$$\sigma_{Y_{lin}^{PT}} = \sqrt{\langle (Y_{lin}^{PT} - \bar{Y}_{lin}^{PT})^2 \rangle_\phi} = \sqrt{2N\left(1 - \frac{v^2}{2}\sin^2\psi\right)} \quad (12b)$$

From Eqs. (12a) and (12b), we see that the quadrature signal uncertainties scale as $\sqrt{N}$, as expected. Furthermore, the above uncertainties alternately vary between maximum and minimum values as a function of the phase $\psi$. The interdependence between the quadrature uncertainties is reflected by the covariance term

$$\text{cov}(X_{lin}^{PT}, Y_{lin}^{PT}) = \langle (Y_{lin}^{PT} - \bar{Y}_{lin}^{PT})(X_{lin}^{PT} - \bar{X}_{lin}^{PT}) \rangle_\phi = -Nv^2 \sin\psi \cos\psi \quad (13)$$

We note that it is not possible to write analytical expressions for the uncertainty or SNR of the complex-valued signal $Z_{lin}^{PT} = |Z_{lin}^{PT}|e^{-i\psi^{PT}}$ in terms of the visibility and phase, given of our assumed form of the probability distribution Eq. (8). However, the uncertainties of $|Z_{lin}^{PT}|$ and $\psi^{PT}$ may be estimated by numerical simulation methods, as we discuss further in the next section.

We define the SNRs of the two quadrature signals as the ratios of the mean absolute value signal $|\bar{Z}_{lin}^{PT}| = Nv$ to the standard deviations:

$$\text{SNR}_X = \frac{Nv}{\sigma_{X_{lin}^{PT}}} = v\sqrt{\frac{N}{2\left(1 - \frac{v^2}{2}\cos^2\psi\right)}} \quad (14a)$$

$$\text{SNR}_Y = \frac{Nv}{\sigma_{Y_{lin}^{PT}}} = v\sqrt{\frac{N}{2\left(1 - \frac{v^2}{2}\sin^2\psi\right)}} \tag{14b}$$

We also define the rate normalized SEs:

$$\text{SE}_X = \frac{\sigma_{X_{lin}^{PT}}}{N} = \sqrt{\frac{2}{N}\left(1 - \frac{v^2}{2}\cos^2\psi\right)} \tag{15a}$$

$$\text{SE}_Y = \frac{\sigma_{Y_{lin}^{PT}}}{N} = \sqrt{\frac{2}{N}\left(1 - \frac{v^2}{2}\sin^2\psi\right)} \tag{15b}$$

In Fig. 4, we show calculations of the standard deviations [Eq. (12)] and the SNRs [Eq. (14)] of the quadrature signals as a function of the parameters $\psi$, $v$, and $N$. Figure 3A shows a schematic of the sampled phase distribution in the complex plane and its relationship to the distributions of the $X_{lin}^{PT}$ and $Y_{lin}^{PT}$ signals, which label the horizontal (real) and vertical (imaginary) axes, respectively. As the signal phase $\psi$ approaches zero ($\pi/2$), the uncertainty in the $X_{lin}^{PT}$- ($Y_{lin}^{PT}$-) quadrature signal approaches its minimum value (= $\sqrt{2N[1-(v^2/2)]}$), while the uncertainty in the $Y_{lin}^{PT}$- ($X_{lin}^{PT}$-) quadrature signal approaches its maximum value (= $\sqrt{2N}$). We note that in the ideal case of $v = 1$, the maximum and minimum uncertainties differ by only a factor of $\sqrt{2}$. In Fig. 4B, we plot the SNR for the $X_{lin}^{PT}$-quadrature signal as a function of the phase $\psi$ for $N$ = 1,000, and $v$ = 1.0, 0.75 and 0.5.

The SNRs shown in Fig. 4B vary sinusoidally with phase between maximum and minimum values ($v \cdot \sqrt{N/[2(1-v^2/2)]}$ and $v \cdot \sqrt{N/2}$, respectively) with contrast that diminishes rapidly with decreasing visibility. The dashed gray curves in Fig. 4B represent the analytical expression given by Eq. (14a), and the colored curves are the results of numerical simulations, which we discuss in the following section.

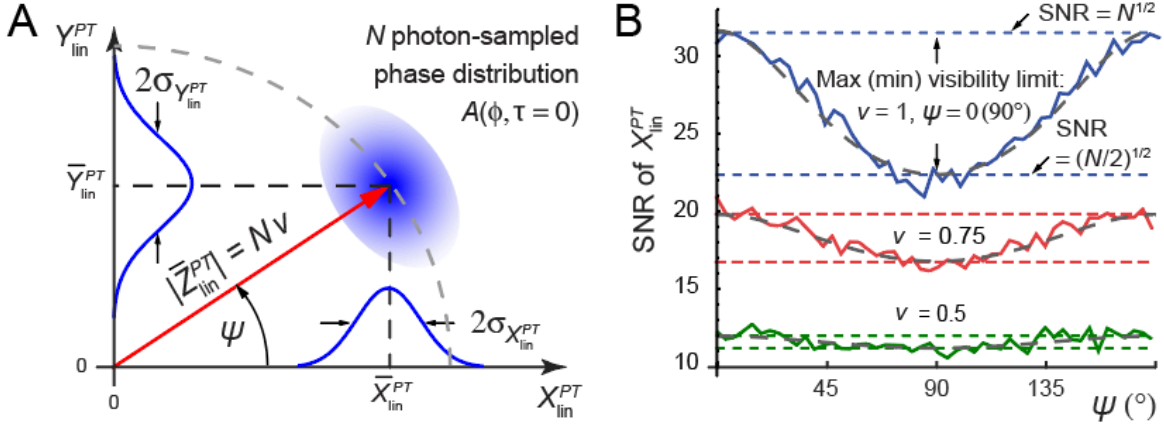

**Figure 4.** (*A*) Schematic of sampled phase distribution for *N* photon events detected within an integration time window $T_{PT}$. The mean signal has phase $\psi$ and amplitude $Nv$, and mean in-phase and in-quadrature signal components $\bar{X}_{\text{lin}}^{PT}$ and $\bar{Y}_{\text{lin}}^{PT}$, respectively. The uncertainties of the quadrature signals [Eqs. (12a) and (12b)] are related to the projections of the sampled phase distribution onto the *X*- and *Y*-coordinate axes. (*B*) Comparison between numerical simulations and analytical expression for the SNR of the $X_{\text{lin}}^{PT}$-quadrature signal [Eq. (14a)] for *N* = 1,000, as a function of the signal phase $\psi$ and visibilities $v$ = 1.0, 0.75 and 0.5.

## 2.5 Numerical simulations

To understand the effects of statistical sampling error on our PTPC measurements, we performed numerical simulations of the linear signal as a function of the flux and visibility. We used statistical sampling methods to generate model signals for the case in which instrument error (e.g. mechanical instabilities, electronic noise, etc.) is negligible. Thus, our numerical simulations isolated the effects of finite sampling of the signal rate on the measurement uncertainties.

In Fig. 5, we illustrate the various steps that we carried out to perform our simulations. First, we determined the laser-molecule overlap spectrum $\hat{Z}_{\text{lin}}(\omega)$ between that of our broadband laser source $|\alpha(\omega)|^2$ and the absorbance spectrum $\hat{\chi}^{(+)}(\omega)$ of the Cy3 dimer DNA system [see Eq. (7)]. The overlap spectrum contains contributions from both the 0-0 and 1-0 vibronic bands of the Cy3 dimer DNA system, as shown in Fig. 5*A*. From the overlap spectrum, we determined by inverse Fourier transform [Eq. (6a)] the expected complex-valued linear signal interferogram $Z_{\text{lin}}(\tau)$, which is down-sampled at the monochromator reference frequency $\omega_R$ (see Fig. 5*B*). The signal interferogram provides the amplitude $|Z_{\text{lin}}(\tau)| = fv(\tau)$ and phase $\psi(\tau)$ as a function of the MZI delay $\tau$. From the amplitude and phase, we reconstructed the high-flux normalized photon rate $A(\phi, \tau)/f$ given by Eq. (2b). In Fig. 5C, we plot $A(\phi, \tau)/f$ for the values of the interferometer delay $\tau$ = 0, 5.3, 10.7 and 16.0 fs. We see that for increasing values of $\tau$, the visibility decreases while the phase shifts toward increasingly negative values.

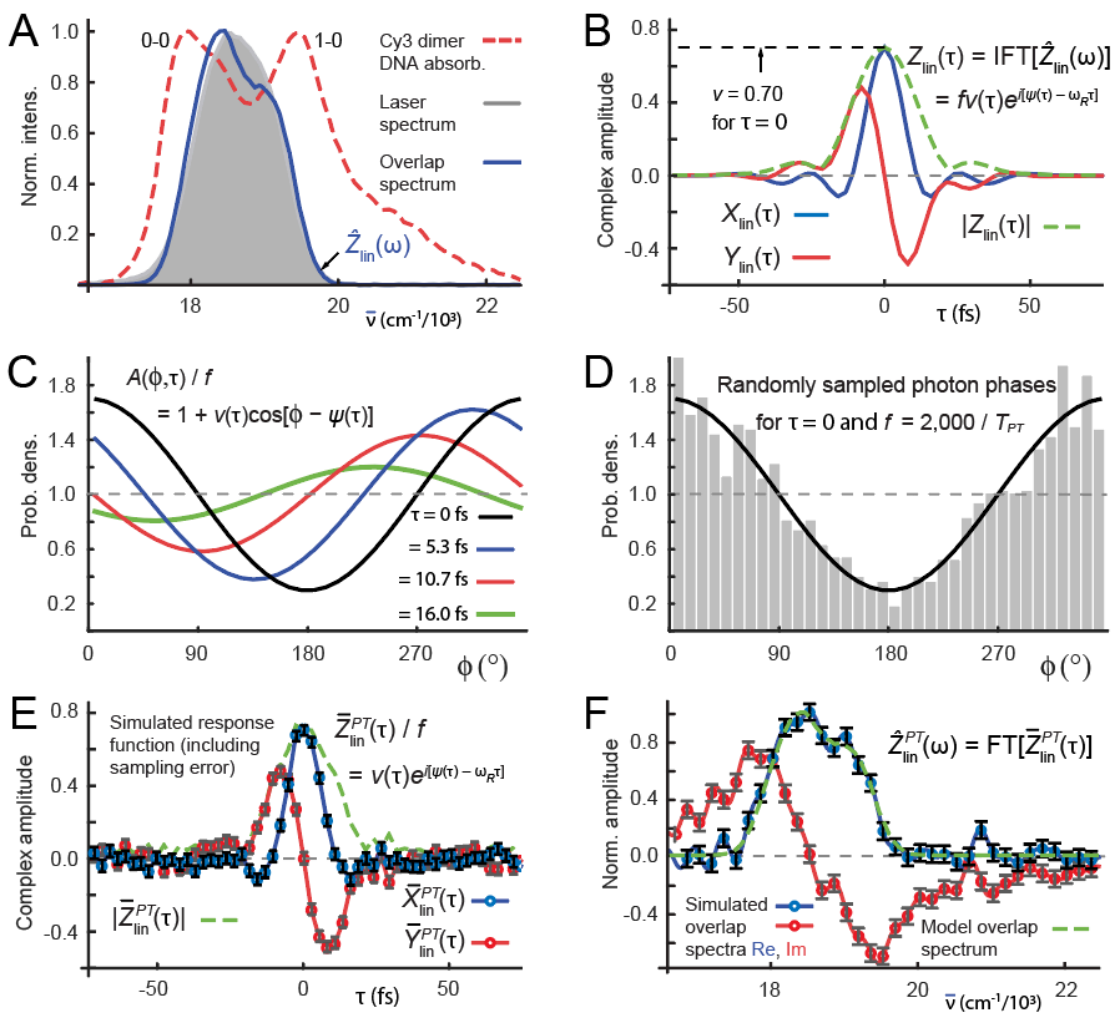

**Figure 5.** Numerical simulations of PTPC experiments. (*A*) Overlay of experimental absorbance spectrum (dashed red), laser spectrum (solid gray) and laser-molecule overlap spectrum (blue). (*B*) The complex-valued signal interferogram with real (blue), imaginary (red) and absolute value (dashed green) parts. (*C*) The expected high-flux normalized photon count rate $A(\phi, \tau)/f$ for values of MZI delay $\tau$ = 0 (black), 5.3 fs (blue), 10.7 fs (red) and 16.0 fs (green). (*D*) Phase histogram of 2,000 randomly sampled photon events during a time window $T_{PT}$ = 1 s and interferometer delay $\tau$ = 0. (*E*) Reconstructed signal interferograms from simulated phase-tagged photon distribution, with real (blue), imaginary (red) and absolute value (dashed green) parts. (*F*) Overlay of theoretical (dashed green) and reconstructed real (blue) and imaginary (red) laser-molecule overlap spectra. In both panels *E* and *F*, the error bars (uncertainties) were determined using the bootstrapping method described in the text.

We used the normalized photon rate Eq. (8) as the probability distribution function from which we applied a numerical random sampling method [al90] to generate a set of $n$ photons with phase assignments within the interval $[0, 360°]$. We based the precise number of events $n$ used for a given data set on a Poisson distribution $P(n) = \bar{N}^n e^{-\bar{N}}/n!$, with $\bar{N}$ the average number of photons detected within the interval $T_{PT}$. In Fig. 5D, we show an example of a simulated phase-tagged photon rate $A^{PT}(\phi, \tau)$ for $\tau = 0$ fs and $f^{PT} = \bar{N}/T_{PT} = 2{,}000/T_{PT}$. From the simulated discrete rate $A^{PT}(\phi, \tau)$, we calculated the mean signal interferogram $\bar{Z}_{\text{lin}}^{PT}(\tau)$ by summing the individual photon phase factors according to Eq. (9b). In Fig. 5E, we show the simulated complex-valued down-sampled interferogram. We determined error bars for each value of the interferometer delay using the bootstrapping method, as described in the previous section. We note that the magnitude of the error is approximately constant over the full temporal range of the scan. The presence of 'noise' in the simulated interferogram is due entirely to the finite statistical sampling of the photon event distribution. From the simulated interferogram we determined the overlap spectrum $\hat{Z}_{\text{lin}}^{PT}(\omega)$ by performing the Fourier transform given by Eq. (6a). As shown in Fig. 5F, the real part of the simulated overlap spectrum aligns closely with the model overlap spectrum, while also accounting for the effects of statistical sampling error. The error bars shown in Fig. 5F were generated using the bootstrapping method, and the magnitude of the error is approximately constant over the frequency range shown.

To summarize this section, performing the numerical simulations described above allows us to examine directly the role of statistical error associated with the finite number of samples of the phase-dependent photon detection rate. By comparing the performance of our PTPC measurements to the results of our numerical simulations, we may assess the importance of instrument noise in our measurements. As stated previously, and discussed further below, we find that statistical sampling error under low-flux conditions is the dominant contribution to measurement uncertainty of our PTPC experiments.

## 3. Results and discussion

We performed low-flux experiments using the PTPC and lock-in detection methods, for comparison. We used both methods to study the same sample as a function of signal flux and visibility. For each value of the MZI delay $\tau$, we detected single photons during an integration time window $T_{PT} = 1$ s. In Fig. 6, we present the resulting $X$- and $Y$-quadrature response functions (panels $A - D$) and their associated laser-molecular overlap spectra (panels $E - H$), measured under two different signal conditions: $\bar{f}^{PT} = 1{,}943$ s$^{-1}$ and $\bar{v}^{PT} = 0.70$ (panels $A$, $B$, $E$ and $F$) versus $\bar{f}^{PT} = 133$ s$^{-1}$ and $\bar{v}^{PT} = 0.56$ (panels $C$, $D$, $G$ and $H$). Measurements carried out using the PTPC method (panels $A$, $C$, $E$ and $G$) are shown in the left-hand column, and those using lock-in detection (panels $A$, $D$, $F$ and $H$) are shown in the right-hand column.

Example histograms of the phase-dependent count rate for these two signal conditions (with $\tau = 0$) were shown earlier in Figs. 3$B$ and 3$D$. We determined the mean in-phase $\bar{X}_{\text{lin}}^{PT}(\tau)$ and in-quadrature $\bar{Y}_{\text{lin}}^{PT}(\tau)$ signals using Eq. (10) (see Figs. 6$A$ and 6$C$), from which we constructed the complex-valued mean signal interferogram $\bar{Z}_{\text{lin}}^{PT}(\tau) = \bar{X}_{\text{lin}}^{PT}(\tau) - i\bar{Y}_{\text{lin}}^{PT}(\tau)$. We also show the error bars ($2\sigma_{X_{\text{lin}}^{PT}}$ and $2\sigma_{Y_{\text{lin}}^{PT}}$) for the $X_{\text{lin}}^{PT}$- and $Y_{\text{lin}}^{PT}$-quadrature components at $\tau = 0$, and we list the SNR $\tau = 0$ in each figure panel for the $X_{\text{lin}}^{PT}$-quadrature signal, which we determined by bootstrapping methods (as described in Sec. 2.4). In general, we found that the SNRs we obtained using the phase-tagging method are very close to the values predicted by Eqs. (14a) and (14b).

In Fig. 6$B$ and 6$D$, we present the lock-in amplifier-detected response functions, which correspond to the $\bar{f}^{PT} = 1{,}943$ s$^{-1}$ and $\bar{f}^{PT} = 133$ s$^{-1}$ signal condition, respectively. For these latter measurements, the TTL pulse stream was sent to the lock-in amplifier (see Fig. 1, red detail) and the $X_{\text{lin}}(\tau)$ and $Y_{\text{lin}}(\tau)$ quadrature signals [given by Eqs. (3c) and (4c)] were measured for each interferometer delay using the lock-in time constant $T_{LI} = 0.1$ s and filter slope set to 24 dB/octave, such that the settling time $10\, T_{LI} = T_{PT} = 1$ s. For the $\bar{f}^{PT} = 1{,}943$ s$^{-1}$ signal condition, the PTPC and lock-in detection methods both provide response functions with similar shapes (compare Fig. 6$A$ to 6$B$). The SNR resulting from the PTPC method is $\sim 2$ times greater than those we obtained using the lock-in detection method. For the $\bar{f}^{PT} = 133$ s$^{-1}$ signal condition, both PTPC and lock-in interferograms appear to be significantly noisy (compare Fig. 6$C$ to 6$D$). Nevertheless, the SNRs resulting from the PTPC method are slightly improved (by a factor of $\sim 1.5$) in comparison to the lock-in detection method.

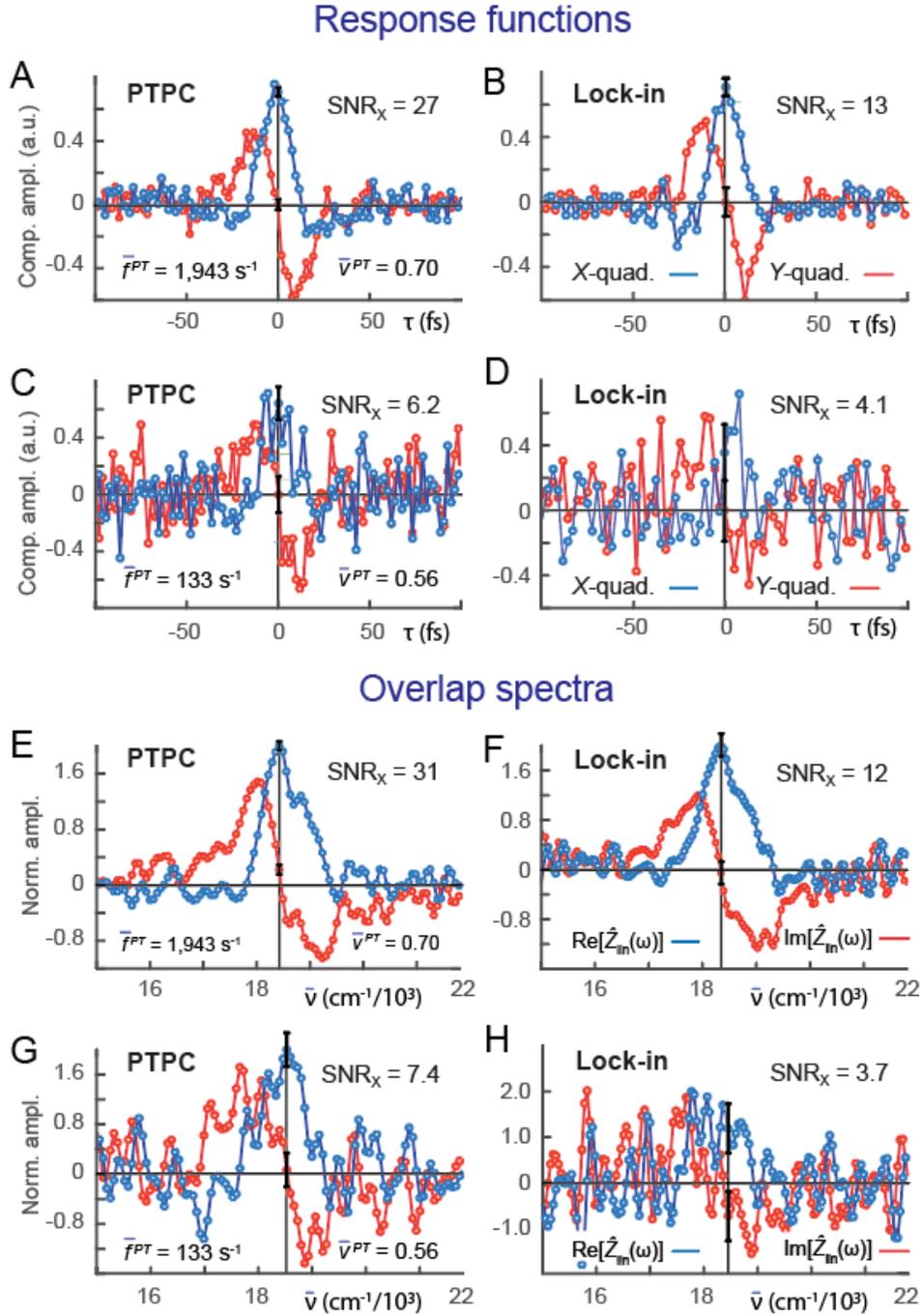

**Figure 6.** Examples of low-flux measurements performed by PTPC (panels *A*, *C*, *E* and *G*) and lock-in detection methods (panels *B*, *D*, *F* and *H*) for various event count rates, measured for one second per MZI delay-stage position. *X*- and *Y*-quadrature response functions are shown (panels *A* – *D*) in comparison to laser-molecule overlap spectra (panels *E* – *H*). Results for experiments with flux $\bar{f}^{PT} = 1{,}943$ s$^{-1}$ and visibility $\bar{v}^{PT} = 0.70$ (panels *A*, *B*, *E* and *F*) are compared to those with flux $\bar{f}^{PT} = 133$ s$^{-1}$ and visibility $\bar{v}^{PT} = 0.56$ (panels *C*, *D*, *G* and *H*).

For each of the response functions presented in Fig. 6A – 6D, we show the corresponding laser-molecular overlap spectrum $\hat{Z}_{\text{lin}}^{PT}(\omega)$ in Fig. 6E – 6H, which was calculated by Fourier transformation with respect to the interferometer delay $\tau$ according to Eq. (6a). For the spectra based on our PTPC measurements, we calculated the error bars and SNRs using bootstrapping methods. For the lock-in detected measurements, we determined the error bars and SNRs in the time domain according to the procedure outlined in Sec. 2.4. To determine the error bars in the frequency domain, we added Gaussian distributed error (assuming the same magnitude as measured as for $\tau = 0$) to the measured response functions at each delay $\tau$, followed by Fourier transformation to obtain a sample spectrum. This procedure was repeated multiple times, and the resulting spectra were used to calculate the standard deviation at each frequency point. In general, the uncertainties we determined for the overlap spectra were similar in magnitude to those of their associated response functions and showed no appreciable covariance across frequencies.

For the $\bar{f}^{PT} = 1{,}943$ s$^{-1}$ signal condition, the overlap spectra determined by both the PTPC and lock-in detection methods exhibited similar shapes (compare Figs. 6E to 6F). In both cases, the real and imaginary parts of the overlap spectra appear to contain contributions from the 0-0 and 1-0 vibronic transitions of the Cy3 dimer optical lineshape (see Fig. 5A), as expected (see Fig. 5F). Notably, the measured SNRs of the PTPC method are 2.5 – 3.5 times greater than those of the lock-in method. For the $\bar{f}^{PT} = 133$ s$^{-1}$ signal condition, the overlap spectra determined by both methods exhibit SNRs < 10, such that the optical lineshapes are difficult to perceive above the baseline signal (compare Figs. 6G to 6H). Even for this ultralow-flux condition, the SNRs of the PTPC method remain approximately twice the value of those produced by the lock-in method.

We next considered the dependence of the uncertainties of the PTPC measurements on the parameters that specify the photon rate probability distribution [Eq. (8)], which are the flux $f$, visibility $v$, and phase $\psi$. We used the statistical sampling calculations described in Sec. 2.5 to simulate the SNRs of the $X_{\text{lin}}^{PT}$-quadrature, the absolute value signal $|Z_{\text{lin}}^{PT}|$, and the rate-normalized SE of the signal phase $\psi^{PT}$ (see Fig. 7). For these simulations, we set the interferometer delay to $\tau = 0$ (corresponding to $\psi = 0$) to calculate the above quantities as a function of the number of photon events $N$ (detected during the time window $T_{PT}$) and the visibility $v$.

In Fig. 7*A*, we present numerical simulations of the $X_{\text{lin}}^{PT}$-quadrature SNR for the range of visibilities $v$ = 0.05, 0.10, 0.22, 0.46 and 1.0. The simulated SNR is in excellent agreement with Eq. (14a), which increases as $\sqrt{N}$ in accordance with the central limit theorem. The $X_{\text{lin}}^{PT}$-quadrature SNR depends linearly on the visibility (to first order). The value SNR = 10 (indicated by the horizontal dashed line) marks an arbitrary boundary, above which the signal can be clearly distinguished from statistical noise. For example, in the limit of ideal visibility $v \approx 1.0$, approximately 200 photons must be included in the integrated signal to achieve an SNR $\approx$ 10, while for the moderate visibility $v \approx 0.5$, approximately 2,000 photons are needed to obtain an equivalent SNR.

In Fig. 7*B*, we plot the results of numerical simulations of the SNR for the absolute value signal $|Z_{\text{lin}}^{PT}|$ as a function of $N$ and $v$. As mentioned previously, it is not possible to write an analytical expression for the uncertainty of the complex-valued signal $Z_{\text{lin}}^{PT} = |Z_{\text{lin}}^{PT}|e^{-i\psi^{PT}}$ in terms of the visibility and phase, as we did for the $X_{\text{lin}}^{PT}$- and $Y_{\text{lin}}^{PT}$-quadratures [Eqs. (14a) and (14b), respectively]. Nevertheless, our numerical simulations show that the SNR of the absolute value $|Z_{\text{lin}}^{PT}|$ increases as $\sqrt{N}$. For the highest visibility $v = 1.0$, approximately 400 photon events are needed to achieve an SNR $\approx$ 10, while for $v \approx 0.5$ approximately 3,000 photons are needed to obtain an equivalent SNR.

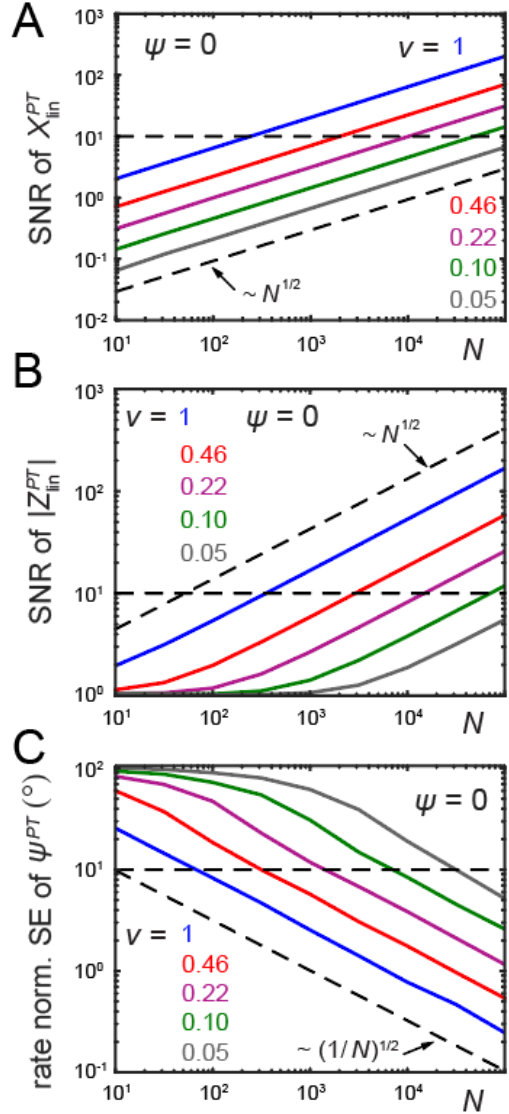

**Figure 7.** Numerical simulations of (*A*) the SNR of the $X_{\text{lin}}^{PT}$-quadrature signal, (*B*) the SNR of the absolute value $|Z_{\text{lin}}^{PT}|$, and (*C*) the rate-normalized SE of the signal phase $\psi^{PT}$ versus the number of detected photons $N$ during a measurement time window $T_{PT}$. For these calculations, the interferometer delay was taken to be $\tau$ = 0 corresponding to $\psi$ = 0. In each panel, the black dashed diagonal lines indicate the expected scaling of the signal metric with detected photon number $N$.

In Fig. 7C, we present the rate-normalized SE of the signal phase $\psi^{PT}$ in units of degrees. The SE decreases as $1/\sqrt{N}$, as expected according to the central limit theorem. The horizontal dashed line indicates an arbitrary set value SE = 10°. This threshold may be reached with approximately 70 photons in the case of ideal visibility, $v = 1.0$, and with approximately 300 photons in the case of moderate visibility, $v \approx 0.5$.

We next compared the dependence of the measurements on the number of photon events using the PTPC and lock-in detection methods. In Fig. 8, we present our experimental results for the SNRs of the $X_{\text{lin}}^{PT}$- and $Y_{\text{lin}}^{PT}$-quadratures (panels A and B, respectively), the absolute value signal $|Z_{\text{lin}}^{PT}|$ (panel C), and the rate-normalized SE of the signal phase $\psi^{PT}$ (panel D). For all these measurements, we set the interferometer delay to $\tau = 0$ (corresponding to $\psi = 0$). We renormalized our data by plotting the ratio SNR / $v$ so that we may compare various measurements performed at different visibility levels. The SNRs and SEs were evaluated using the bootstrapping and other statistical procedures described in Sec. 2.4.

In each of the panels shown in Fig. 8, we compare our PTPC measurements using an integration time window of $T_{PT} = 1$ s (shown as red points) to two sets of lock-in detection measurements using different instrument settings. In the first set of lock-in measurements (blue points), the lock-in time constant was set to $T_{LI} = 0.1$ s and the filter slope to 24 dB/octave, such that the measurement settling time is $10T_{LI} = 1$ s. The detected photon number $N$ was thus adjusted by varying the signal flux. In the second set of lock-in measurements (green points), the signal flux was held fixed to 315 s$^{-1}$, and the lock-in time constant was adjusted to control the photon number.

Both PTPC and lock-in detection measurements appear to obey the expected $\sqrt{N/2}$ and $\sqrt{2/N}$ scaling laws for the SNRs [Eq. (14)] and the SE [Eq. (15)], respectively. Furthermore, the PTPC method outperforms the lock-in method by a factor of ~ 2 over the full range of flux values investigated. We note that for the $X_{\text{lin}}^{PT}$- and $Y_{\text{lin}}^{PT}$-quadrature signals (Figs. 8A and 8B, respectively), the SNRs of the PTPC measurements (red points) follow closely the theoretically predicted values [Eqs. (14a) and (14b), respectively, shown as dashed diagonal lines]. In particular, the agreement for the $Y_{\text{lin}}^{PT}$-quadrature signal appears to agree exactly with the value $\sqrt{N/2}$, as described by Eq. (14b) with $\psi = 0$. The excellent agreement between experiment and theory – over the full range of flux values investigated – suggests that uncontrolled sources of instrument noise (e.g. mechanical vibrations, detector shot noise, etc.) are essentially absent from our PTPC measurements.

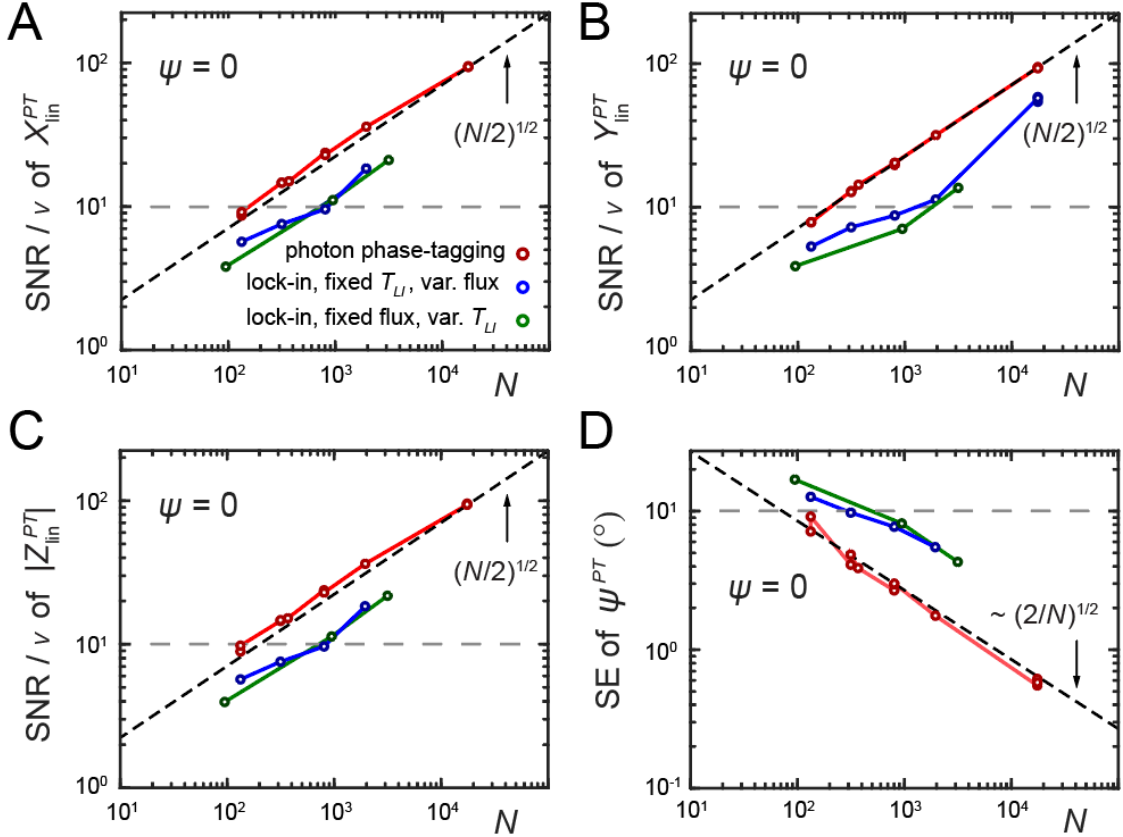

**Figure 8.** Comparison between measurements performed by PTPC and lock-in detection methods. The interferometer delay was set to $\tau = 0$ (corresponding to $\psi = 0$). Visibility-scaled SNRs (i.e. SNR/$v$) are plotted versus detected photon number $N$ for (**A**) the $X_{lin}^{PT}$-quadrature signal, (**B**) the $Y_{lin}^{PT}$-quadrature signal, and (**C**) the absolute value signal $|Z_{lin}^{PT}|$. (**D**) The rate-normalized SE is plotted for the signal phase $\psi^{PT}$ in units of degrees. Measurements performed by the PTPC method (red points) used an integration time window of $T_{PT} = 1$ s. Lock-in detection measurements were performed by varying the flux while using a fixed time constant $T_{LI} = 0.1$ s and filter slope of 24 dB/octave, such that the measurement settling time was equal to 1 s (blue points). Lock-in measurements were also performed using a constant mean flux of $\bar{f}^{PT} = 315$ s$^{-1}$ while systematically varying the time window (green points). The colored line segments are 'guides for the eye.' The horizontal dashed lines indicate the arbitrary threshold of SNR = 10 in panels **A** – **C**, and SE = 10° in panel **D**.

## 4. Conclusions

In this work, we have demonstrated a photon counting phase-tagging (PTPC) approach to Fourier transform (FT) spectroscopy that is well suited to experiments carried out under low-flux conditions. In the low-flux regime, the signal is an intermittent stream of single photon events with the probability to detect an individual photon given by the mean (or high-flux limiting) signal rate. For experiments carried out under moderate-to-low-flux conditions, a lock-in amplifier can be employed with a sufficiently long integration time window to obtain a time-averaged photon count rate. Nevertheless, the PTPC method outperforms lock-in detection under these conditions because it uses all of the information available from individual photon events. PTPC thereby determines the signal mean and standard deviation without loss of measurement precision.

We performed test experiments on a model molecular system for which the phase-swept fluorescence signal flux and visibility could be systematically adjusted, and a direct comparison could be made between PTPC and conventional lock-in detection. In addition, we performed numerical simulations to examine the role of statistical error associated with the finite number of samples of the phase-dependent photon detection rate. The quadrature signals were determined by using a fast-digital electronic counter to assign an interferometer phase to individual photon detection events. We evaluated the signal performance in terms of the signal-to-noise ratio (SNR) and the rate-normalized standard error (SE).

For a range of signal flux and visibility levels, we find that the PTPC approach outperforms conventional lock-in detection by a factor of ~ 2 or greater. Moreover, measurement uncertainties are dominated by statistical noise at low-flux levels and are not affected by mechanical instabilities of the interferometer or other forms of instrument noise. Our studies establish the range of signal parameters needed to perform FT spectroscopy experiments on molecular systems under low-flux conditions. For example, at relatively high visibility levels (i.e. $v \approx 1.0$), approximately 200 photons must be detected within a time window $T_{PT}$ to achieve a SNR $\approx 10$ for the $X$-quadrature signal, while for the moderate visibility $v \approx 0.5$, approximately 2,000 photons are needed to obtain an equivalent SNR. The value SNR = 10 delineates an arbitrary boundary above which the signal can be unambiguously distinguished from statistical error.

The PTPC method presents a number of opportunities for carrying out FT spectroscopy experiments in which the signal flux is expected to be low. By combining time- and phase-tagging of the detected photon data stream, it is possible to extract linear and nonlinear (higher harmonic) signal components in post data acquisition, and to study the fluctuations of these signals from non-stationary molecular systems. For example, low-flux FT spectroscopy may be performed on single-molecule emitters to enable time-dependent studies of the molecular optical response and laser-molecule overlap spectrum. Higher-order signal contributions at multiples of the modulation frequency, such as two-photon absorption [ka16, br15a, br17], can be isolated in principle from the photon signal stream. A straightforward extension of the current setup is to include a second interferometer to excite samples using four pulses to perform fluorescence-detected two-dimensional FT spectroscopy [te07, ae11, gr17, ti18]. In such experiments, simultaneous dual phase-tagging of the photon signal stream will enable the acquisition of nonlinear response functions and two-dimensional optical spectra. Another potential application is to monitor molecular systems excited by low-flux time-frequency entangled photon pairs (or EPP) for which weak signals and long integration times are anticipated [ra13, la20, ta20]. The benchmark studies of the current work establish the range of signal parameters over which future low-flux experiments may be attempted.

## Acknowledgements

We thank Boulat Bash for helpful discussions on error estimation. We also thank Prof. Brian Smith for useful discussions.

## Funding

This work was supported by grants from the John Templeton Foundation (RQ-35859 to A.H.M. M.G.R. and B.S. as co-PIs), by the National Science Foundation Chemistry of Life Processes Program (CHE-1608915 to A.H.M.), and by the National Science Foundation RAISE-TAQS Program (PHY-1839216 to A.H.M. M.G.R. and B.S. as co-PIs).

## Disclosures

The authors declare no conflicts of interest.